# Fuzzy based Edge Enhancement using Wavelets and Energy Operator

S.Anand, G. Sangeetha Priya
Department of ECE
Mepco Schlenk Engineering College, Sivakasi, India, 626005

*Abstract*

**Edge enhancement and preservation of edges to emphasize the features for images is an essential task in computer vision. The conventional operators may cause false edge detection. In this paper, a fuzzy inference system (FIS) is proposed, which combines with Wavelet (WT) based multidirectional and multilevel edge finding technique for getting good edge localization. After applying of WT decomposition on the original image, the coefficients are filtered out from horizontal, vertical, and diagonal directions. A non-linear energy operator is applied to all WT decomposition to corroborate the edge pixels. This energy calculated edge coefficients are given into the FIS for better enhancement. Since WTs allow for separation of the image into frequency bands without affecting spatial locality, the real edge pixels are further emphasized.**

**Indexing Terms: Edge detection, wavelet transform, the fuzzy inference system**

## I. Introduction:

Detection of edges in an image is one of the most important steps in a complete image understanding system [1], [2], [3], [4]. To achieve satisfactory results, it is relevant to select object contours that are significant with respect to human perception, without being deceived by the noise which could be exist in the data. Different approaches have been used to solve this problem. The best known of which is the evaluation of an estimate of the local gradient or of the local second derivative. Few fuzzy techniques have especially addressed the problem of edge detection [5], [6]. A different approach to edge detection using a family of new rule based operator proposed by many papers. These operators are based on IF-THEN-ELSE architecture which makes them able to perform many important image processing tasks [7]. The overall method of the proposed approach is given in Fig. 1.

In this respect, this paper proposed an edge detector different from previous work which combines the three WT image coefficients along with FIS to detect the edges. This paper is organized as follows section II introduces discrete WT; section III presents the simplest energy operator; section IV describes the fuzzy inference system; Section V illustrates the results and finally, this paper reports the conclusions.

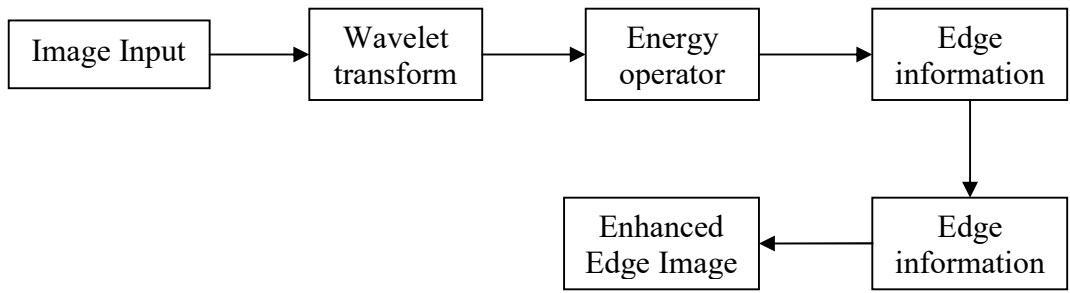

Fig. 1 Block diagram of the proposed method

## II. Discrete Wavelet Transform

WTs are functions generated from one single function ψ by translations and dilations. The principle behind the WT is to characterize any functions as combinations of many scaled and dilated WT functions. Any such decomposition decomposes the given 1D or 2D functions into different scale levels, whether each level is further decomposed with a resolution adapted to that level.

There are two types of WT decomposition for two-dimensional image data namely Discrete WT and Stationary WT. Stationary WT normally finds its applications in image texture analysis and enhancement techniques. Decompositions of an image using the Discrete WT are depicted in Fig.2.

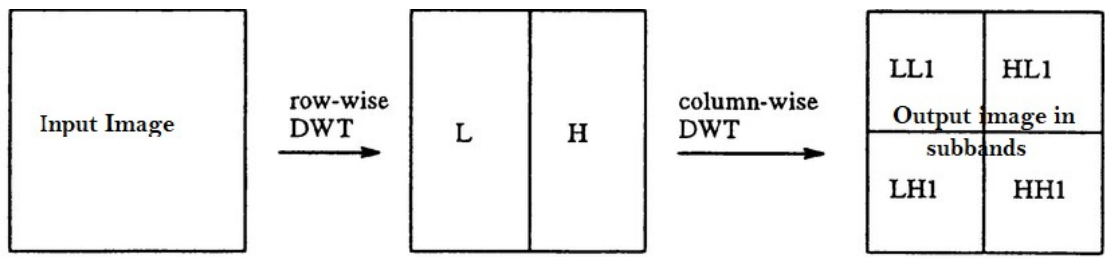

Fig.2 2D Discrete Wavelet transform for two dimensional signals

The filters H and G are one dimensional high pass and low pass filter respectively for image decomposition. The filters are applied both for rows and columns and finally the decomposition provides sub bands corresponding to different resolutions levels and orientation. The decomposed image is reconstructed using a reconstruct filter [5].

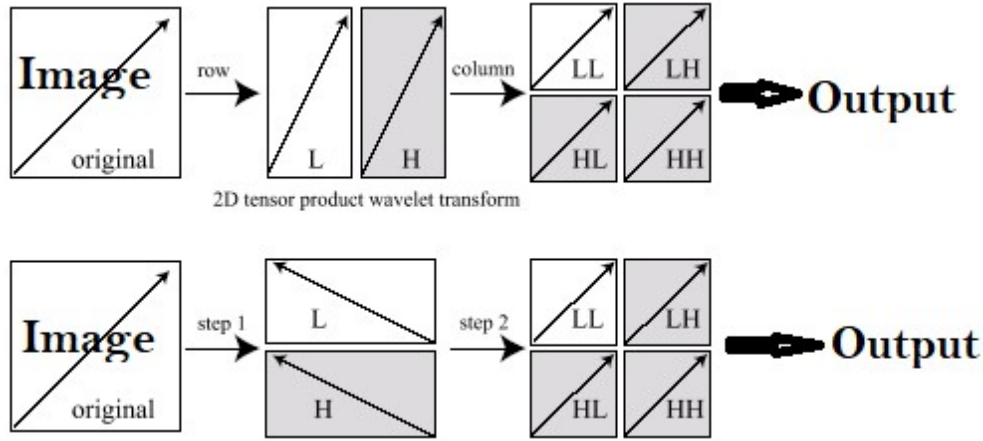

**Fig. 3  2D Discrete Wavelet transform implemented on an image using separable property**

Computation of the corresponding WT amounts to two successive steps, i.e., first filtering the image row by row and then column by column; see Fig.3. Both steps are related to dilation matrices respectively and dilate only in one direction.

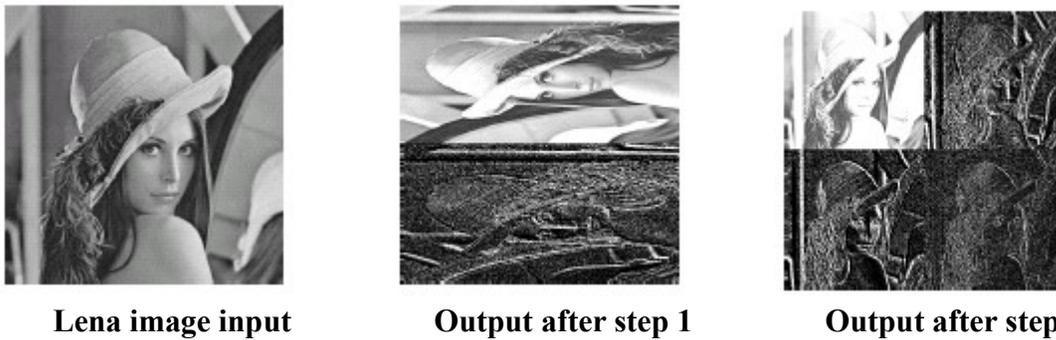

    **Lena image input**    **Output after step 1**    **Output after step 2**

**Fig. 4  2D Discrete Wavelet transform implemented on Lena image using separable property**

The dilation matrices are applied twice to obtain tensor-product-like WTs. In order to get the same sub band decomposition, the dilation matrices generated WT on both the low-pass and the high-pass sub bands are applied. See Fig. 4 for an illustration.

### III. Energy Operator

   The energy operators are usually operated on the image pixels to energize the pixel than the neighboring values. Since the decomposed WT coefficients are small in gray values, the energy operator can be used to corroborate the WT coefficients. The energy operator serves its purpose in two ways [2] i) to detect the edge pixel independent with the background. ii) Energize than the neighboring pixels.

We have chosen the energy operator to compact with the discrete WT coefficients like vertical, horizontal, and diagonal. The generalized monlinear energy operator is given below $T[f(n)] = f^2(n) - f(n+1) * f(n-1)$.

### IV. Fuzzy Inference Systems

**Image Pre-Processing:** During the input image pre-processing stage, the image is WT decomposed. Two intermediate stages (row wise WT operated and column wise WT operated) of WT applied on Lena image shown in Fig.5 and this magnitude is higher for lower frequencies, and that the filters in low pass and lower the magnitude correspond to higher frequencies.

**Fuzzy Sets and Membership Functions:** The system application was supported by the 8 bit input image and the output is attained by after applying of defuzzification process. Both input and outputs are maintained in 8 bit (i.e.) 0 – 255 levels. Their gray levels are among 0 and 255 and these values describe the various intervals of the output and the input. Three fuzzy sets were formed to characterize the input image intensity intervals and they were associated with the linguistic variables known as "high", "medium" and "low". The membership functions for the edge fuzzy sets are described with the input image intensities and the output intensities are 0, 128 and, 255, as shown in Figure 4. They are derived from a standard Gaussian function which existing between '0 to 255'. For all the input images, the linguistic variables "medium" and "low" was chosen (Figure 4).

**Defuzzification and fuzzy logic:** The functions are recognized to tool the minimum 'and / norm-T' functions and maximum 'or / norm-S' functions. The Mamdani method was selected as the defuzzification process. In this procedure, thee fuzzy sets that are obtained by put and processed on each implication rule to the energy operator activated wavelet coefficients of image input were combined through add operational function. The system output pixels for edges were resulted from the output membership function.

**Inference Rules Definitions:** The fuzzy inference rules were defined from the notion that the FIS system output which corresponds to 'edges' is high only for those pixels are being edges in the input image. The robustness of the FIS is described as to brghtness variations (pixel contrast) in the form of variance also considered when these rules were obtained. The rubrics were well-defined to describe the general idea that in pixels going to an edge or not. FIS identify the 'where there is a large gray level variations in the horizontal / vertical direction'. For example, participation image pixels are 'low' then the output (nonedge) variable also 'low' and when all the inputs are 'high' and the outputs (belongs to edge) are also considered to be 'high' etc. All the rules are expressed and inferred by the FIS.

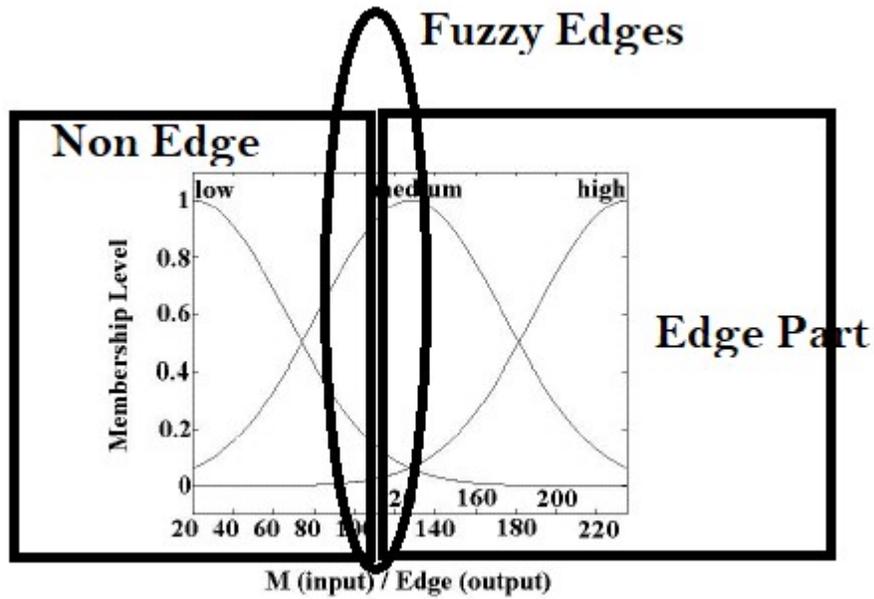

Fig. 5 Membership functions for fuzzy based edge detection

V. Experimental Results

The system was tested with cameraman image, and the performance is compared to edge detection by Sobel and Canny operator. The weights related with each fuzzy rules are adjusted to permit correct outcomes to be found while finding the edges of the image shown. The contrast variations are more in the image region. While performing experimental tests, all parameters were retained constant. It is observed, that the Sobel operator with automatically estimated threshold from the image's root mean squared (RMS) value does not permit edges to be detected in the lower contrast regions Fig. 6. It is also observed that pixels not edges pixels, mostly in the low gray level regions (with low values). The FIS system, detect the edges even in the regions with low contrast. This is due to the improved localization property of the WT while decomposing the image into sub bands [2].

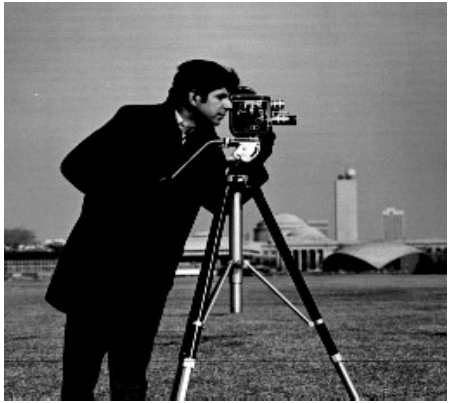

(a) Cameraman image

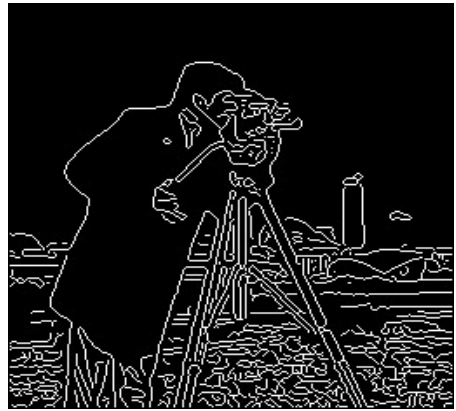

(b) Canny Edges

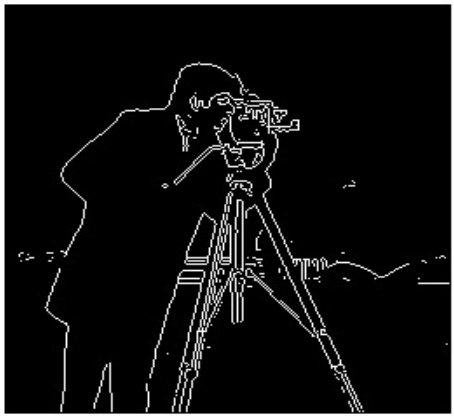

(c) Sobel Operator Edges

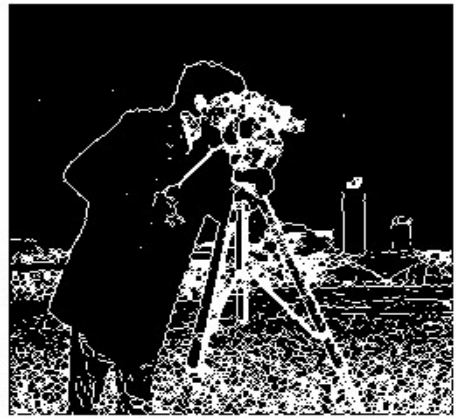

(d) Edges by Proposed method

Fig. 6 Comparison of output images (a) original (b) Canny edge (c) Sobel Edge (d) Proposed method

## VI. Conclusion

An energy operated FIS system for edge enhancement is presented in this paper. From output results of FIS, it is concluded that the proposed method e qualitative better than the Sobel operator and canny operator. The applied FIS system along with nonlinear energy operator offering better robustness to contrast variations, along with good localization with the support of WT in edge detection and enhancement.